# Exciplex-driven blue OLEDs: unlocking multifunctionality applications

Dominik Weber[1,2], Annika Morgenstern[2], Daniel Beer[3], Dietrich R.T. Zahn[2,4], Carsten Deibel[3], Georgeta Salvan[2,4]*, Daniel Schondelmaier[1*]

[1]Nanotechnology, University of Applied Sciences Zwickau, 08056 Zwickau, Germany

[2]Semiconductor Physics, Chemnitz University of Technology, 09107 Chemnitz, Germany

[3]Optics and Photonics of Condensed Matter, Chemnitz University of Technology, 09107 Chemnitz, Germany

[4]Center of Materials, Architectures and Integration of Nanomembranes, Chemnitz University of Technology, 09107 Chemnitz, Germany

**Abstract**

We present the development of multifunctional blue-emission organic light-emitting diodes (OLEDs) using TADF-exciplex materials. These OLEDs exhibit sensitivity to external stimuli and achieve a maximum external quantum efficiency (EQE) of 11.6 % through partly liquid processing. This technique allows for large-scale production on arbitrary geometries.

The potential multifunctionality of the devices arises from their response to low external magnetic fields (up to 100 mT) with an efficiency up to 2.5 % for magnetoconductance, while maximum magneto-electroluminescence effects of 4.1 % were detected. We investigated novel aspects, including the utilization of two organic materials without further doping and the investigation of the impact of 2,2',2"-(1,3,5-Benzinetriyl)-tris(1-phenyl-1-H-benzimidazole) (TPBi) processing in liquid and vapor form. The insights gained provide a fundamental understanding regarding the applicability of exciplex (EX) materials for fully solution-processed OLEDs through a deliberate omission of doping. Our work represents a significant advancement on the path towards multifunctional OLED technology, with potential applications in cost-efficient, scalable organic full-color displays and advanced sensing system

*Corresponding authors:
E-Mail address: daniel.schondelmaier@fh-zwickau.de
                salvan@physik.tu-chemnitz.de

## 1 Introduction

Organic light-emitting diodes (OLEDs) based on reverse intersystem-crossing (RISC) processes such as thermally activated delayed fluorescence (TADF) exciplex couples have great potential in flat-panel displays and solid-state lighting due to their high internal quantum efficiency [1–6]. Additionally, these materials display sensitivity to stimuli originating from their surroundings, particularly magnetic fields, thereby enhancing their appeal for sensor applications [7]. Considerable research efforts are intensively being conducted aiming for higher efficiencies, optimized layer stacks, narrow blue emission and deposition on arbitrary geometries to implement the OLEDs in *e.g.* flat displays [8], flexible electronics [9], and point-of-care devices in medical applications [10]. In 1961, Parker *et al.* [11] first reported the phosphorescence and thermally activated delayed fluorescence of eosin, demonstrating the TADF effect. The work by Endo *et al.* [12] introduced TADF systems with electroluminescence, and in ref. [3], the first system with a high external quantum efficiency (EQE) of 5.3 % was presented, relying on an intrinsic small energy gap between singlet and triplet states ($\Delta E_{ST}$). The latter yields a comparably high rate of RISC between the triplet and singlet states. In general, TADF materials provide a means to increase the efficiency of OLEDs from the theoretically achievable 25 % to 100 % as they allow to harvest the triplet excitons in addition to the singlets ones [13]. Next to exciton-based TADF materials an efficiency of 100 % might be utilized by intermolecular excited states (exciplex states), which also exhibit efficient RISC [1,6,14]. In this case, there is a high probability that a charge transfer state forms at the interface between an electron acceptor and an electron donor. The electron transition occurs between the acceptor LUMO and the donor HOMO to the new exciplex state formed in between [15, 16], with a small singlet-triplet gap of a few meV in the range of thermal energy at room temperature. The first exciplex-based emission was reported by Hu *et al.* in 1994 [17]. The endeavors of the Adachi group were seminal in this field, as they presented pioneering advancements with an inaugural EQE of 5.4 % in 2012 for green-emitting OLEDs. Furthermore, within the same year, they achieved a remarkable EQE of 10 % for the emission of pure blue light [14, 18]. The emission process is very similar to that of exciton-based TADF materials, with the advantage lying in the simple fabrication of exciplex materials, where no mixing or elaborate synthesis methods are required and a wide range of donor and acceptor molecules is available [19]. Additionally, exciplex systems facilitate Förster energy transfer and increase device efficiency while suppressing efficiency roll-off [4, 7, 20]. This efficiency improvement is particularly crucial for blue-emitting OLEDs, which commonly exhibited lower efficiencies and lifetimes compared to green and red-emitting OLEDs [21]. However, achieving precise blue emission is crucial for full-color display applications and white light generation [22,

23]. Unfortunately, the production of blue OLEDs poses several challenges, due to the difficulty of combining a large bandgap and a narrow emission band (high color purity), as well as the limited availability of suitable materials that fulfil the requirements and the complexity of the required layer stacks [21, 24]. Furthermore, OLED fabrication is aimed at applications on arbitrary geometries, where the standard deposition method of organic materials, namely vacuum thermal evaporation, is not suitable. Therefore, solution processed devices come into play, which provide the possibility for scalable and cost-efficient production, but suffer from lower efficiency and shorter device lifetime [25]. Hence, this study focuses on the utilization of the active materials tris(4-carbazoyl-9-ylphenyl)amine (TCTA) and 2,2',2''-(1,3,5-Benzinetriyl)-tris(1-phenyl-1-H-benzimidazole) (TPBi) and poly(N-vinylcarbazole) (PVK) as hole transport layer (HTL). These materials have already been investigated as host material for blue emission OLEDs, *e.g.* Wang *et al.* reported Bis(2,4- difluorophenylpyridinato)- tetrakis(1-pyrazolyl)borate iridium(III) (Fir6) (20 wt%):TCTA with an EQE of 5.73 % at a wavelength of 458 nm [26], while Wu *et al.* studied (acetylacetonato)bis[2-(thieno[3,2-c]pyridin-4-yl)phenyl]iridium(III) (PO- 01):(TCTA:TPBi) (0.03:1 by weight) with an EQE of 14.75 % [4]. However, these studies employed doping agents to enhance stability and efficiency. Additionally, the influence of PVK as HTL for TCTA was already examined, where both materials were directly mixed or spin-coated sequentially on top of each other [27].

To unlock the multifunctionality of our OLED structures, we investigated the light-emitting capability as well as its sensitivity to an external magnetic field [28–30]. In recent years, several studies reported that some OLED architectures, when exposed to external magnetic fields, exhibit a change in conductivity (organic magnetoconductance, OMC), resistivity (organic magnetoresistance OMAR), and electroluminescence (magneto -electroluminescence, MEL). These magnetic field effects (MFE) are typically observed at magnetic fields up to 100 – 200 mT [29, 31]. In general, MFE can be attributed to spin mixing channels [32]. Several theoretical models have been proposed *e.g.*, bipolaron model [33], loosely-bound polaron-polaron pair model [34], triplet-exciton polaron quenching model [35], singlet-triplet interconversion and gyromagnetic ($\Delta g$) mechanism [36–38]. A detailed discussion of these models attempting to describe MFE in organic materials is beyond the scope of this work and can be found, elsewhere [38]. However, the origin of MFE in organics is still a matter of debate [29]. It is expected that more than one mechanism can contribute to MFEs and that, depending on the type of the charge carriers and on the presence of photoexcited carriers, different mechanisms can dominate. Using the approach for radiative recombination through RISC [22, 23], record OMC values above 1000 % at room temperature were reported for an OLED structure based on 4,4',4''-tris[phenyl(m-tolyl)amino]triphenylamine (m-MTDATA) as electron transporting layer (ETL) and tris(2,4,6-trimethyl-3-(pyridin-3-yl)phenyl)borane (3TPYMB) as hole transporting layer (HTL) [7]. The organic MFE strength is highly dependent on the singlet-triplet states gap, temperature, active layer thickness as well as the electrical conditioning[1]. Hence, these results surpass even the record for inorganic MFE, *i.e.* a magnetoresistance value of 604 % reported for magnetic tunnel junctions (MTJs) at room temperature [39].

In this study, we present a comprehensive comparison of deep blue OLEDs based on emission from TCTA and TPBi, and demonstrate that the deep blue emission is possible without the incorporation of additional dopant materials, which distinguishes this work from others [40]. Typical optoelectronic characteristics, as well as the MFE, are compared among different layer stacks and processing methods. The primary emphasis lies in the comparative analysis of fully solution-processed organic layers (PVK, TCTA, and TPBi) and the synergistic approach involving solution processing (PVK and TCTA) combined with vapor deposition (TPBi). It was previously observed that the intermixing of TCTA and TPBi lead to a clear red shift in the emission spectrum compared to the single compounds which can be attributed to the successful intermixing of TCTA and TPBi compounds and the exciplex formation in the system [41]. The integration of liquid processing and vapor deposition techniques, coupled with appropriate material selection, opens possibilities for production of OLEDs on a large scale but also allows for their fabrication through printing techniques in the future. Additionally, applications extending beyond electroluminescent devices, towards advanced magnetic field sensing could be envisaged.

---

[1]*Electrical conditioning is the process of preparing or adjusting an electrical device. Typically, the device is subjected to a controlled electrical load or cycles to improve its stability, reliability, or performance. This process may involve activities such as switching the power supply, applying certain voltage or current levels to achieve the desired properties or behaviors.*

## 2 Materials and Methods

### 2.1 Materials

The substrates used were pre-structured ITO glasses (ITO glass OLED substrates - pixelated anode (6 pixels) from Ossila). The organic materials utilized in the fabrication process included a water-based solution of Poly-(3,4-ethylendioxythiophene)-poly-(styrene sulfonate) (PEDOT:PSS) from Heraeus (Clevios™ - PEDOT:PSS). Additionally, Molybdenum(VI) oxide powder ($MoO_3$), Poly(9-vinylcarbazole) (PVK), and Lithium fluoride (LiF) were purchased from Sigma-Aldrich. Moreover, 2,2',2''-(1,3,5-Benzinetriyl)-tris(1-phenyl-1-H-benzimidazole) (TPBi) and tris(4-carbazoyl-9-ylphenyl)amine (TCTA) were obtained from Ossila. All Materials were used without further purification. The chemical structure as well as the band diagram with the appropriate HOMO-LUMO levels for each molecule can be found in Fig. 1b), a), respectively.

The PEDOT:PSS solution was diluted with isopropanol (IPA) at a ratio of 1:0.04 w%. A mixture of $MoO_3$ powder dispersed in an ammonium hydroxide solution (0.25 g/ ml $NH_3OH$) was added at a mass ratio of 1:0.02. PVK and TCTA were dissolved individually in chlorobenzene (10 mg/ ml and 15 mg/ ml, respectively). TPBi was dissolved in IPA at a ratio of (1.6 mg/ ml). Prior to application, all solutions were treated in an ultrasonic bath for a minimum of 30 min at 80 °C and filtered with a Nylon syringe filter with a pore size of 0.22 µm. The remaining materials were used as received and applied through thermal evaporation.

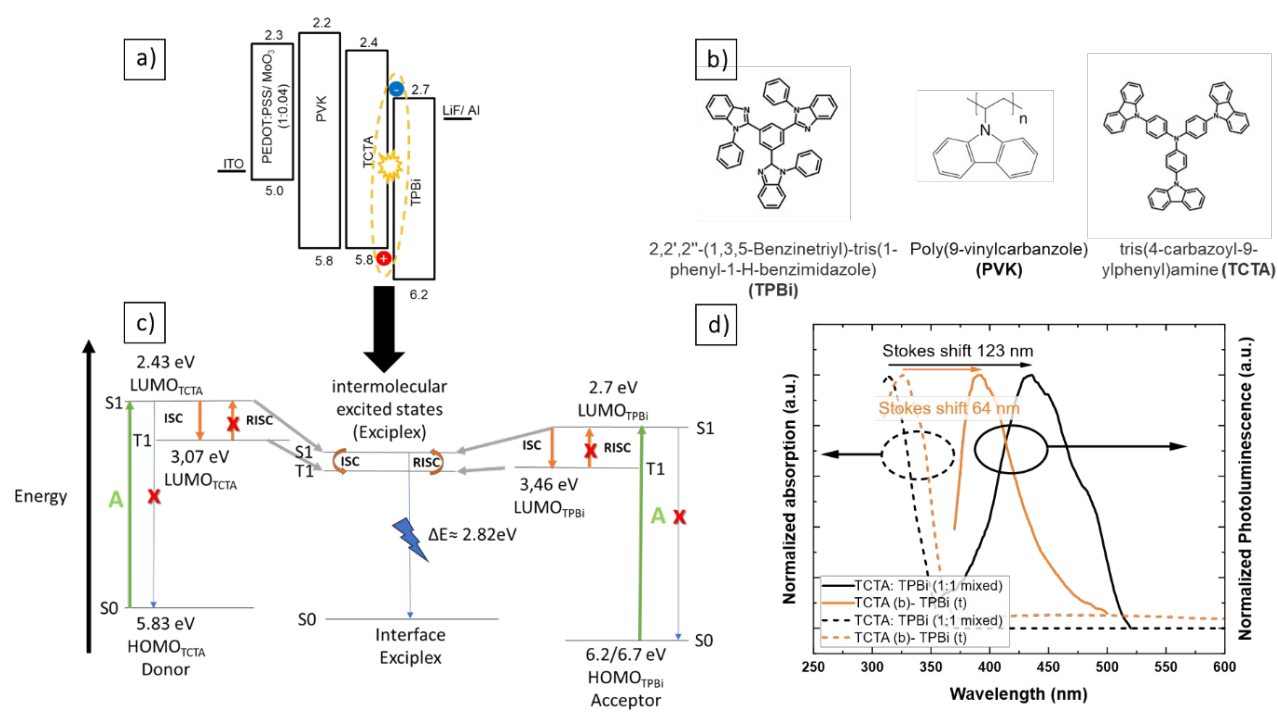

***Fig. 1*** *a) Band diagram for the OLED layer stack used in this work, where an exciplex formation occurs between TCTA and TPBi b) Chemical structures for the active material used in the study, namely TCTA and TPBi as well as PVK (used as HTL). c) Modified diagram representing electronic transitions based on exciplex emission. d) Recorded absorption (dashed lines) and PL spectra (solid lines) for TCTA and TPBi spin-coated separately and in a mixture (1:1).*

**2.2 OLED Fabrication**

The complete production process is sketched in Fig. 2-1, providing an image for each step. The ITO glass substrates, with a thickness of 115 nm (cf. Fig. S2) and a sheet resistance of 20 Ω/ square, underwent a cleaning process involving alkaline solvent and hot IPA treatments, each lasting 20 min at 70 °C, followed by exposure to a UV ozone chamber for 30 min. A 35 nm thick layer of the PEDOT:PSS /MoO$_3$ mixture, serving as the hole transport layer (HTL), was spin-coated in an atmospheric environment at a rotational speed of 5000 rpm and then annealed on a hot plate at 130 °C for 30 min (cf. Fig. 2-1 a). After the PEDOT:PSS layer had completely dried, the following materials (PVK, PVK:TCTA mixture at a 1:1.5 ratio, TCTA, and TPBi) were deposited in a glovebox with oxygen content below 1 ppm and humidity below 1 %. In the spin coating process, 5 drops of the PVK, PVK:TCTA, or TCTA solutions were carefully dispensed onto the substrate at a spin speed of 3000 rpm. After an initial 5 s period, the spin speed was raised to 1000 rpm and maintained for 10 s to ensure a uniform distribution of the solution. To achieve the desired final thickness of 30 nm, the spin speed was then promptly increased to 3000 rpm without any interruption in the operation of the spin coater (cf. Fig. 2-1 b), c)). TPBi was also deposited using this three-step process, with a final spin speed of 2000 rpm to achieve a thickness of 20 nm. All layers were annealed at a temperature of 130 °C for 30 min before applying the next layer (cf. Fig. 2-1 d) (S)). Each layer thickness of the spin-coated layers was determined by profilometry using a Dektak instrument (cf. Fig. S2). The resulting layer stacks are illustrated in Fig. 2-3. The remaining layers, namely TPBi (20 nm, if not already liquid-applied) (cf. Fig. 2-1 d) (E)), LiF (0.7 nm), and aluminium (120 nm), were deposited using thermal evaporation in a vacuum chamber at a pressure of 10$^{-6}$ mbar (cf. Fig. 2-1 e)). The growth rate for the organic materials was maintained at 1 Å/s, while LiF was deposited at 0.1 Å/s and aluminium at 5-10 Å/s (cf. Fig. 2-1 e)).

## 2.3 Measurement details

The photoluminescence spectra (PL) were recorded by a Cary Fluorescence spectrometer, where the excitation wavelength for TCTA, PVK, and TCTA: TPBI was chosen to be $\lambda_{ex}$ = 330 nm and for TPBi $\lambda_{ex}$ = 305 nm. The UV-vis absorption spectra were detected using a Cary 60 UV-vis spectrometer. The baseline was determined using quartz glass and subtracted from the spectra of interest. For the PL and UV-vis measurements, films of the single materials (TCTA, TPBi, PVK) and of the TCTA:TPBi (1:1) mixture as well as the stack of TPBi (t)[2] and TCTA (b)[3] were produced by spin coating. The parameters were chosen according to the ones used for the OLED manufacturing. To perform photometric characterization of the fabricated OLEDs, a Gigahertz-Optik UMBB-210 integrating sphere (sphere diameter of 210 mm) was utilized in conjunction with a calibrated photodiode PD-1101 (measuring range from 250 nm to 1100 nm). The optical radiation dependence on wavelength was evaluated using an Ocean Optics Flame-S-UV-VIS-ES spectrometer.

---

[2] (t)- top layer
[3] (b)- bottom layer

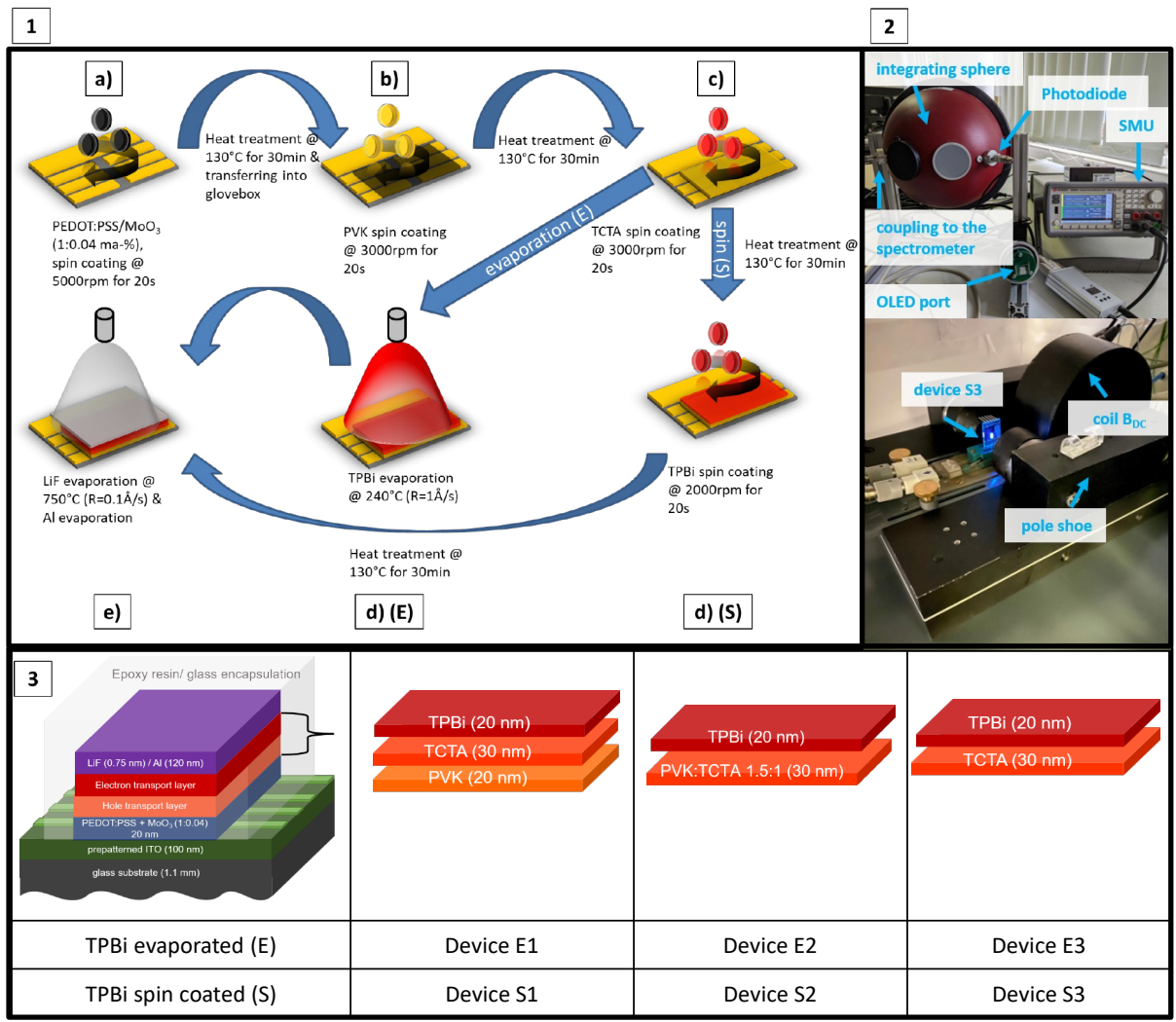

***Fig. 2*** *1 OLED production steps. 2 The measurement setup for the OLED characterization (top) and the home-built measurement setup for ON-OFF magnetoconductance measurements with a device S3 type OLED. 3 Scheme of the layer stack for the different OLEDs produced in this work. Devices, where the TPBi layer was spin-coated, are denoted as S and the ones, where it was evaporated, are denoted as E. The underlying layers are all spin-coated: PVK and TCTA (single layers), PVK: TCTA (1:1.5) (mixed layer) or TCTA (from left to right, respectively).*

The OLEDs voltage supply and photodiode current measurement were accomplished using the Keysight B2902A Source Measurement Unit (cf. measurement setup in Fig. 2-2). The wavelength, diode characteristics, and photodiode current provide the necessary parameters for the calculations of the current density, luminance, and EQE.

These parameters enable the determination of key metrics related to the OLEDs performance and efficiency. The calculation of the EQE from the measured luminance is based on the standard model [42].

For the influence of the external magnetic DC field onto the conductivity of the OLEDs, a homebuilt setup was used (cf. Fig. 2-2). The current through the OLED was measured by a Keithley Source meter (SMU) 2636, while a constant voltage was applied to the device. Additionally, an out-of-plane DC magnetic field was applied to the OLED. The magnetic fields of 0 mT, 100 mT, 0 mT, and – 100 mT were alternately applied (20 data points for each cycle). The fields, were generated by a current through the coil of $I_B$ = 2 A, which corresponds to $B_{DC}$ = 100 mT (or $I_B$ = 0 A, $B_{DC}$ = 0 mT; $I_B$ = -2 A, $B_{DC}$ = -100 mT). The current was generated by an EA-PSU 9080-50 power supply. The spectra were recorded by a Stellar Net Black Comet spectrometer, where a fiber was placed directly in front of the emissive area of the device.

## 3 Results

### 3.1 Photoluminescence and Absorption

The OLED layer stack is designed to allow the formation of exciplexes through the intermixing of the TCTA and TPBi compounds. According to the $EX_{Triplet}$-$EX_{Singlet}$ gap between the two materials, disregarding the exciton binding energy, an emission peak at ~ 2.8 eV is expected for a TCTA:TPBi exciplex [43]. The shallow HOMO level of TCTA, which arises from aromatic amine units, facilitates the formation of an exciplex when combined with the electron transporting material TPBi, which can be ascribed as dimerization [44]. PVK is utilized in the OLED layer stack as a second HTL in between PEDOT:PSS/ $MoO_3$ and TCTA. PVK can influence the hole mobilities, which change the charge carrier balance and therefore enhances the device stability and influences the EQE [45, 46]. PL spectroscopy was used to investigate the exciplex formation in these mixed active materials, comparing them to the neat ones. Fig. S 1 demonstrates that TPBi (spin coated and evaporated), TCTA, and PVK exhibit emission maxima at 3.3 eV, 3.19 eV, and 3.04 eV, respectively.

However, to comprehend the effect of spin-coating the TPBi layer on top of the on TCTA, we conducted PL spectra and UV-vis analysis on the mixed TPBi and TCTA sample (1:1 mixture) and compared them (see Fig. 1 d) to the neat ones (cf. Fig. S1). When TPBi (t) was deposited onto pre-deposited TCTA (b), the main emission peak remained almost in the same position as pristine TCTA, with a slight red shift of 2 nm. The shift in the main emission peaks can be correlated with a change in the electronic structure, as not all peaks are preserved [47]. In the case of a 1:1 mixture of TPBi and TCTA, there was a pronounced red shift in the emission to a peak wavelength of 436 nm. The red shift when mixing TCTA and TPBi (1:1 mixture) as well as a broadening towards higher wavelength for spin-coating TPBi (t) on top of TCTA (b) indicate a subtle mixture at the interface. Interfacial mixing through spin-coating with non-orthogonal solvents was also reported in other investigations [48]. A singlet energy for the exciplex of (2.81 ± 0.10 eV), consistent with the literature [44], was determined, while the absorption was detected at an energy of (3.96 ± 0.20 eV). The difference between the PL spectra and the maxima (optical gap) originates from accessible exciplex states below the band gap [49]. Please note that the value of the Stokes shift (1.15 eV) between the absorption and PL (Fig. 1 d)) can be influenced by the relative position of the donor–acceptor units. An additional influence can be caused by polar solvents, which further increase the Stokes shift [50].

**3.2 Optoelectronic properties**

Fig. 3a) exemplarily shows the electroluminescence (EL) spectra of the OLED stacks investigated in this work. The EL of the OLEDs, in which TPBi was spin-coated exhibit a red shift compared to the case, when TPBi was evaporated. This red shift indicates an increased exciplex formation within the system, which is achieved by spin coating [6]. By selecting IPA as the solvent for TPBi, the detachment of the underlying layer can be markedly reduced, as evidenced by measurements using a profilometer and the nearly constant thickness of the TCTA layer (cf. Fig S2). This is attributed to the low solubility of TCTA in IPA. In Fig. 3 b), most of the OLEDs show a common diode characteristic, besides OLEDs S1 and S3 where a high leakage current is particularly noticeable (it can be seen by the substantial current density at negative voltage). These high leakage currents suggest that spin coating can lead to an elevated degree of impurities, or damages in the OLED layers such as pinholes [51–54]. The highest voltages applied at the end of the voltage range, as observed in Fig. 3b), are characterised by an exponential increase in current density, which can result in the OLEDs short. In Fig. 3 c), the luminance is plotted as a function of voltage, providing a comparison of the light emission behaviour of the OLEDs. It clearly shows the significantly higher quality of the evaporated OLEDs E1, E2, and E3 compared to the spin-coated ones, which exhibit lower luminance. However, it is important to note that device E2 exhibits a significantly lower turn-on voltage, more than 2 V lower compared to devices E1 or E3. When compared to a reference study [55] that extensively investigated the impact of blending PVK and TCTA, it is observed that the hole mobility of a TCTA:PVK mixture, starting from pristine PVK with a mobility of $1.98 \times 10^{-6}$ cm$^2$ (Vs)$^{-1}$, decreases to $3.86 \times 10^{-7}$ cm$^2$ (Vs)$^{-1}$ when 20 % TCTA is added to PVK. However, it rises to $4.89 \times 10^{-4}$ cm$^2$ (Vs)$^{-1}$ when the TCTA concentration exceeds 50 %. This suggests that TCTA has a higher hole mobility, while the injection barrier from TCTA to PEDOT:PSS is lower. Consequently, this is the reason for the reduction of the turn-on voltage.

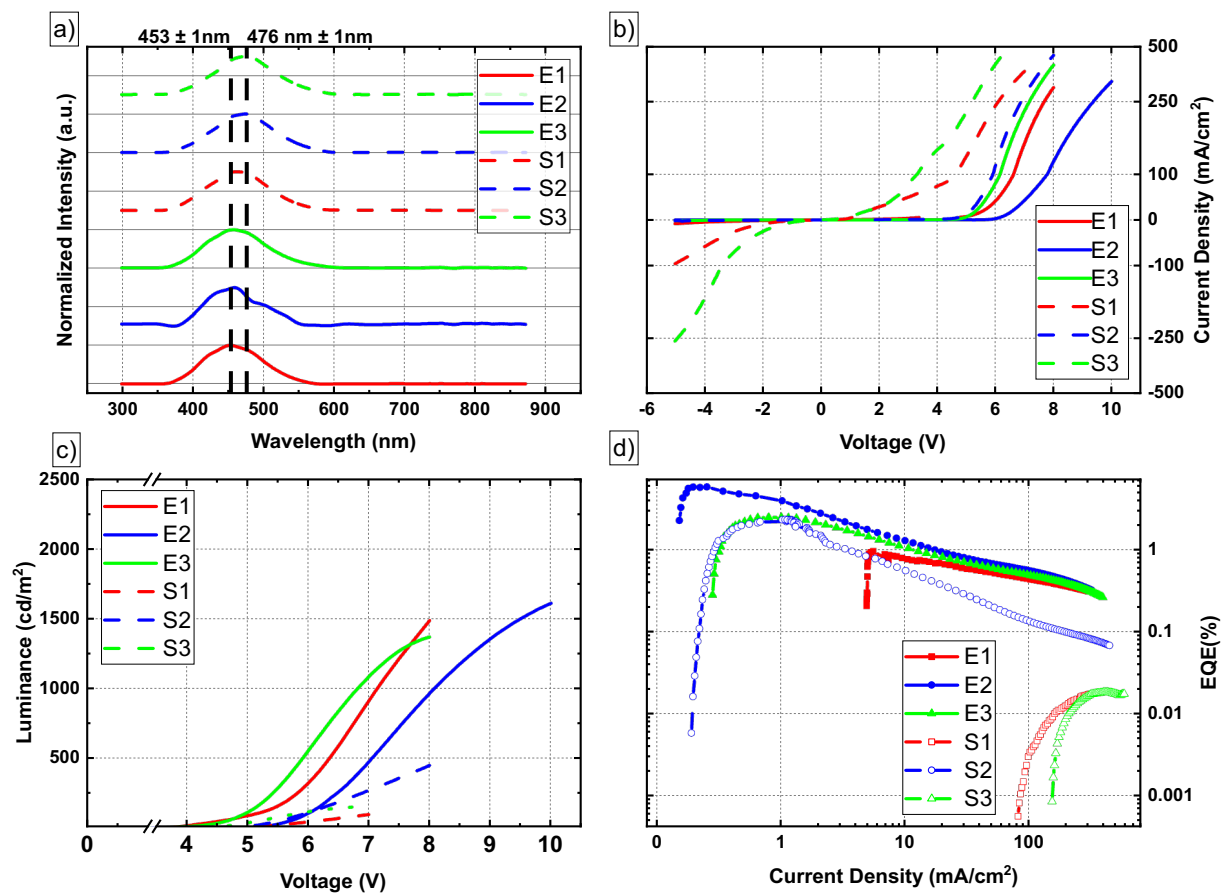

***Fig. 3*** *Optoelectronic properties of devices fabricated in this work (see Fig. 2-3. for the device labels): a) Normalized intensity distribution of the electroluminescence as a function of wavelength; b) Current density as a function of applied voltage; c) Luminance as a function of voltage; d) EQE as a function of current density.*

In the same reference study [55], which utilized the red emitter material Ir(piq)2(acac), it was found that the ideal mixing concentration for TCTA in PVK is 5 %. Furthermore, the authors noted that the optimal mixing concentration depends not only on improved mobilities and injection barriers but also on the specific layer stack used. It is worth mentioning, that the choice of mixture concentration between TCTA and PVK compounds can be further investigated. Nevertheless, all layer stacks containing PVK revealed significantly higher EQE values. Fig. 3d) presents the EQE at the corresponding luminance levels. The highest quantum efficiency was observed in OLED E2, reaching 11.6 %. This is remarkably high for an undoped system, as the typical quantum efficiency for similar systems is around 1 % [45]. Furthermore, the EQE of OLED S2 exhibits a substantial increase in comparison to the other spin-coated OLEDs, reaching a value of 0.26 % compared to only 0.02 % in OLEDs S1 and S3. A summary of the maximum performance parameters of each OLED is presented in Table 1.

*Table 1.* *Maximum performances of the devices 1-3 (E and S), respectively. The turn-on voltage $V_{on}$ (at 10 cd/m$^2$), emission wavelength $\lambda$, and EQE are shown below and were extracted from the measurements depicted in Fig. 3.*

| Device | Max performance | | | | | | | | |
|---|---|---|---|---|---|---|---|---|---|
| | $V_{on}$ (V) | | | $\lambda$ (nm) | | | EQE (%) | | |
| | 1 | 2 | 3 | 1 | 2 | 3 | 1 | 2 | 3 |
| TPBi evaporated (E) | 4.6±0.2 | 2.1±0.1 | 4.7±0.1 | 453.6±1.0 | 459.7±1.0 | 457.9±1.0 | 0.4±0.2 | 11.6±0.4 | 0.6±0.2 |
| TPBi spin-coated (S) | 5.3±0.3 | 4,7±0.3 | 4.8±0.4 | 461.4±1.0 | 474.8±1.0 | 476.2±1.0 | 0.02±0.01 | 0.26±0.05 | 0.01±0.01 |

### 3.3 Organic magnetic field effects

Notably, we observed discernible changes in the response to the alternating switching on and off the DC magnetic field, with distinct increases (or decreases) of the current density.

The OMC(B) is defined in Eq. (1) according to ref. [56]:

$$OMC(B) = \frac{I(B) - I(0)}{I(0)} \tag{1}$$

with I(B) and I(0) corresponding to the current in the presence and absence of the external magnetic field, respectively.

Fig. 4 illustrates the results of OMC for the various devices (cf. Fig. 2-3). From the measured data the slope of I(t) was subtracted, to emphasize the influence of the magnetic field on the measured current. The raw data can be found in Fig. S 3.

All material combinations, except device S3, exhibited a significant increase in conductivity when the magnetic field was applied. The observed OMC effect shows no dependence on the direction of the magnetic field *e.g.*, regardless of whether the magnetic field was positive or negative, we observed a comparable shift in current density. The time-dependent OMC curves can also serve to estimate the system's response time. In our investigation, this response time was found to be instantaneous within the time scale of the measurements, which falls within the range of seconds. This fast response time is not typical for all organic compounds *e.g.,* TIPS Pentacene as an example of a material with longer response time [57]. Moreover, devices S1 and S2 showed higher OMC values compared to their half-evaporated counterparts. Besides a few works, *e.g.* [15, 58], this is an instance of such effects observed in solution-processed devices, indicating its uniqueness in the field. Furthermore, we found that devices with higher EQE revealed lower MFE (cf. EQE and OMC of S 2). The increased MFE reveal a stronger intermixing of the organic layers, thus leading to an increased contact area between TCTA and TPBi, which raises the possibility for exciplex formation leading to higher MFE effects.

By comparing the devices with and without PVK, it is apparent that the OMC does not correlate with its presence, since all devices (with and without PVK) reveal noticeable OMC effects. Nevertheless, PVK increases the device stability and hence the degradation during the measurement cycles is decreased (cf. Fig. 4 OMC of device E2 and E3 and Fig. S3). The OMC$_{max}$ values are summarized in Table S2. The highest value was observed for device S2 with a OMC$_{max}$ value of (2.54 ± 0.57) %. Comparable values in the range of a few percent have been found for other TADF- exciplex systems with the active materials m- MTDATA and B4PYMPM [41].

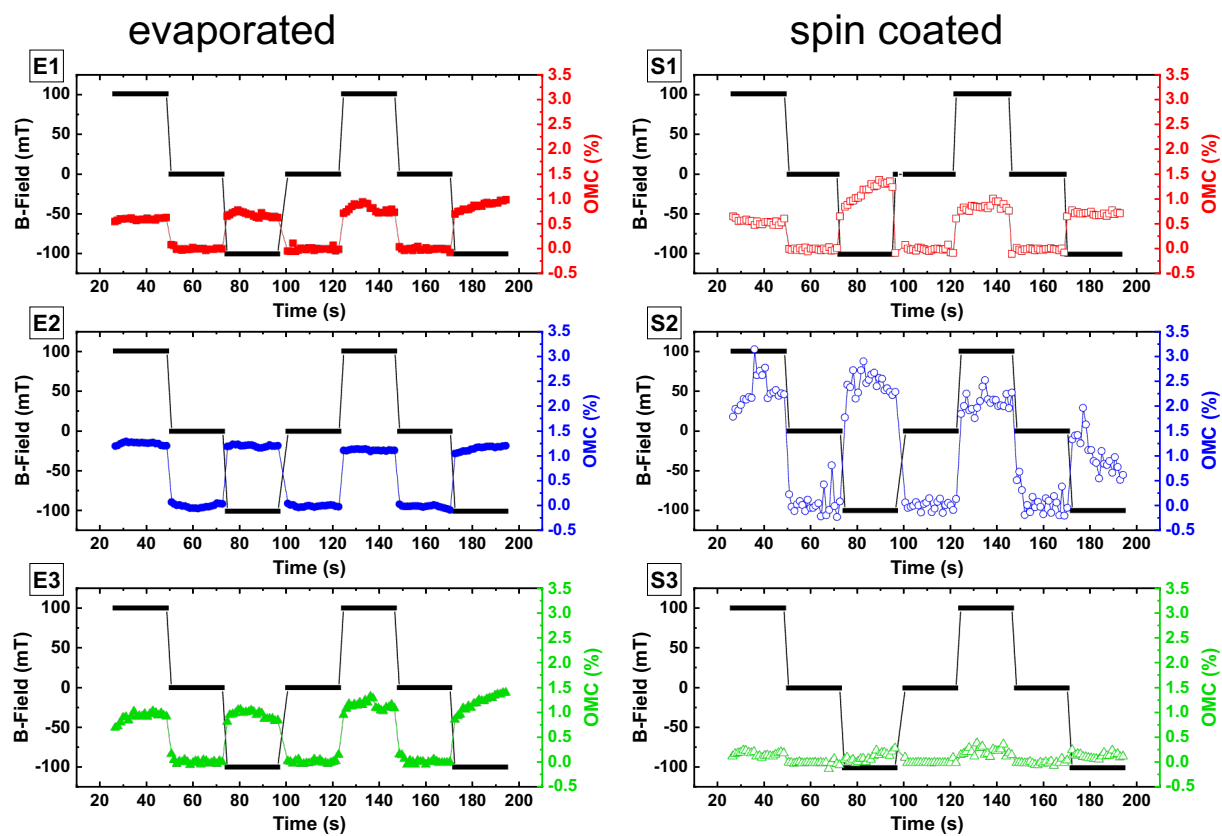

**Fig. 4** *The variation for the OMC for the different OLEDs, where E and S correspond to the layer stacks described in Fig. 2- 3. The OMC is displayed as a function of the magnetic field during the cycles of on (B-field = 100mT), off (B-field = 0mT), and on (B-field = -100mT).*

We also conclude, that PVK exhibits no sensitivity to external magnetic fields. The low OMC effects for device S3 can be related to the low device stability, due to the missing hole transport layer. One possible way of increasing the MFE in all devices, including S3, might be electrical conditioning, as previously suggested for other material combinations [7, 20]. In addition to the OMC investigations, we monitored the MEL by measuring the spectra for each on-off cycle within the first five seconds. In Fig. 5, we exemplarily depict the results for devices S2 and E2. The corresponding uncorrected MEL cycles to the OMC results displayed in Fig. 4 can be found in Fig. S4. In the case of device E2, the current exhibited more stability throughout the measurement, which is evident from the shallower slope observed (cf. maxima taken at $B_{on}$ and $B_{off}$, of OMC in Fig. 5). Hence, the MEL intensity of device E2 remained more stable. This demonstrates the enhanced stability achieved when TPBi was deposited by thermal evaporation.

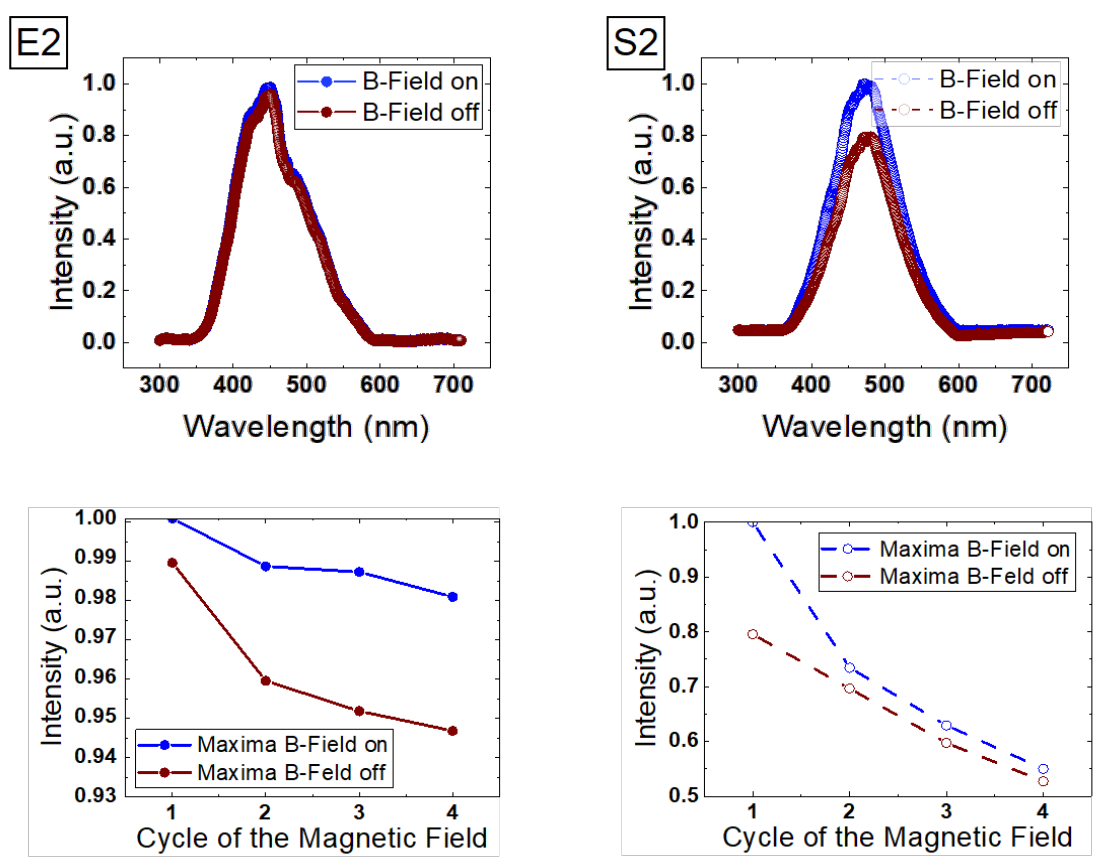

*Fig. 5* In the top row the spectra recorded at the second on-off (device E2, (device S2) cycle (cf. Fig. 4) are shown. The bottom row shows the maximum electroluminesczens intensity of each magnetic field cycle. The external magnetic field was alternately switched on and off ($B_{on}$ = 100 mT).

However, it is important to note that for both devices, a clear difference in intensity was observed when comparing the maxima. Both, the peak position and the full width at half maximum (FWHM) remained constant during the entire measurement for both devices (cf. Table S1). This indicates that while the intensity varied, the spectral characteristics of the emitted light remained unchanged. The occurrence of the MFE can primarily be ascribed to alterations in population resulting from spin mixing when an external magnetic field is applied, namely hyperfine interaction [15]. Further, for TADF-exciplex materials the Δg mechanism has to be considered, where the RISC is modulated by an external magnetic field. This additional spin mixing channel is available according to the different chemical environment of the electron and hole forming the exciplex [15, 32, 59], which may lead to a dephasing of the charge carrier species. To detect this phenomenon more closely, one would need to sweep the magnetic field and would expect a decreased curve after $MFE_{max}$ is reached. This work will be part of a further study.

## 4  Discussion and Conclusion

In this study, we focused on the development of fully-solution processed and partly solution processed / partly evaporated OLED devices along with their sensitivity to external stimuli and production methods. Through photoluminescence spectroscopy, we obtained compelling evidence of exciplex formation in the mixed materials TCTA and TPBi. The fully liquid-processed devices (except S3) exhibited higher MFE (compared to their half-evaporated counterparts), indicating a higher degree of intermixing between the two exciplex materials. The noticeable red shift in the electroluminescence spectra observed for the fully spin-coated OLEDs , as compared to evaporated devices, serves as an additional indication of exciplex formation and improved intermixing. The polymer PVK plays a significant role as the second hole-transporting material in the device, particularly when combined with TCTA. The hole mobility decreases due to the presence of PVK, leading to a better balance of charge carriers in the emissive layer. Consequently, this results in increased efficiency, the maximum achievable luminance, and device stability. The intermixing of PVK and TCTA results in an EQE of up to 11.6 % for vapor-deposited TPBi OLEDs and up to 0.3 % for fully solution-processed OLEDs. It is noteworthy that both of these values signify a tenfold increase in efficiency when compared to OLEDs lacking a blended PVK layer or entirely devoid of PVK manufactured using an identical process. This underscores a critical parameter for further investigations, emphasizing the importance of optimizing efficiency through the careful selection of the PVK and TCTA blending ratio. Furthermore, it is noteworthy that the system exhibits the highest MFE if the PVK: TCTA blend is used as donor. In the case of the vapor-deposited TPBi layer, there is a 1.3 % increase in conductivity at $B_{ext}$ = 100 mT, while a 2.5 % increase was observed in the case of fully liquid-processed OLEDs and even 4.1 % for the magneto electroluminescence response.

These results demonstrate, on one hand, the feasibility of OLED systems based on blue emitter materials completely deposited from solution, and on the other hand, the significantly higher sensitivity of solution-processed OLEDs to external stimuli related to the enlarged TCTA /TPBi partial mixing. Density functional theory calculations would be required to shed light on the dimerization of TCTA and TPBi molecules and the TADF process. The emergence of MFE in TADF-exciplex materials is attributed to the interplay of hyperfine interaction and the dephasing of electron and hole precession frequencies (around the external magnetic field) and requires further studies for a deeper understanding of critical aspects such as polaron-pair lifetime and time dependency of the MFE. Nevertheless, our results pave the way towards prospective applications, including multifunctional OLEDs integrated into arbitrary geometries, advancements in magnetic sensing applications, and the development of spintronic devices.

## Conflict of Interest

The authors declare no conflict of interest.

## Acknowledgment

We would like to acknowledge Prof. Michael Mehring (Coordination Chemistry) and especially Adrien Schäfer for the possibility to use their UV-vis spectrometer. Furthermore, Dominik Weber, Annika Morgenstern and Georgeta Salvan would like to thank the SAB for funding this research under the project number: 100649391 (RESIDA).

## Data Availability Statement

The data that support the findings of this study are available from the corresponding author upon reasonable request.

## Authors contributions

D. Weber, A. Morgenstern, G. Salvan and D. Schondelmaier conceptualized the research work. D. Weber and A. Morgenstern have equally contributed to the manuscript. D. Weber and A. Morgenstern prepared the samples. D. Weber performed the optoelectrical measurements and analyzed the data. A. Morgenstern performed the MFE and UV-vis measurements and analyzed the data. D. Beer and A. Morgenstern performed the PL measurements and A. Morgenstern analyzed the data. D. Weber and A. Morgenstern wrote the original version of the manuscript. C. Deibel, D.R.T. Zahn, G. Salvan, D. Schondelmaier validated the data. G. Salvan and D. Schondelmaier supervised the work. All authors reviewed and edited the

# Supplementary

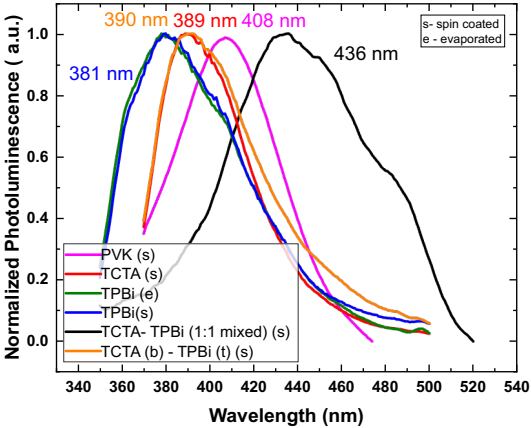

***Fig. S 1*** *Photoluminescence spectra for thin films of the organic compounds used in the OLED fabrication, namely TCTA, TPBi (evaporated (e) and spin coated (s)), PVK, and TCTA:TPBi (1:1) mixture as well as the layers TCTA (b)[1] and TPBi (t) deposited on each other (trace the layer intermixing of the devices).*

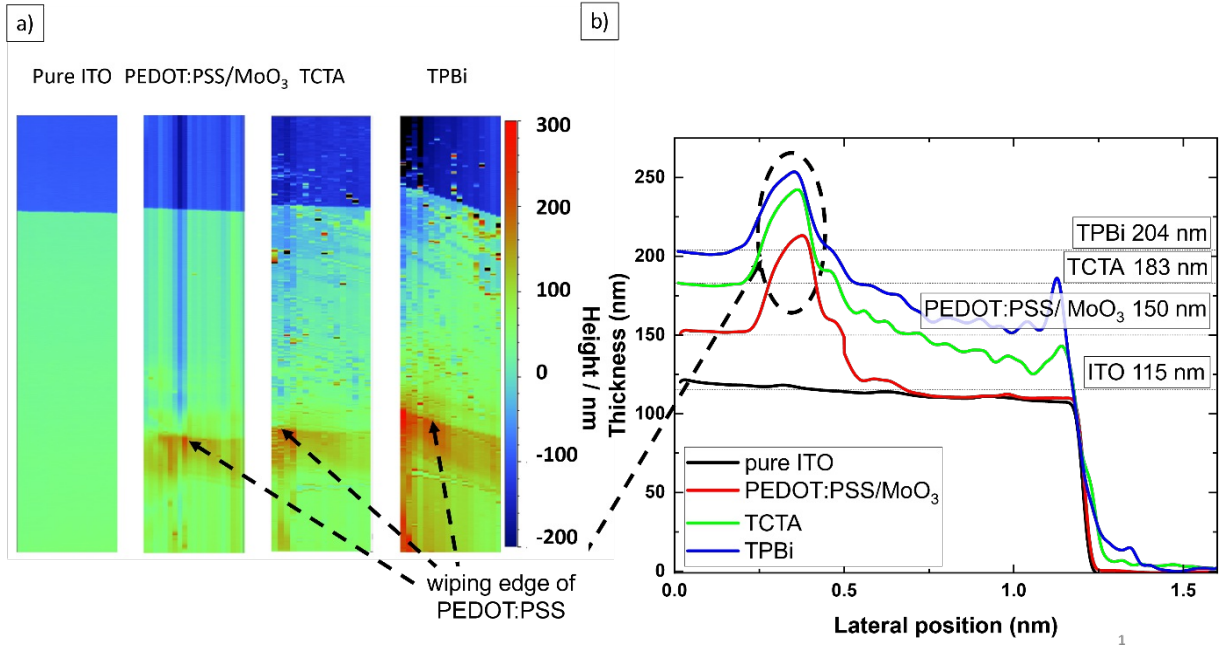

***Fig. S 2*** *Exemplary contour plots for Sample S2 (a) and the corresponding height profile (b). The measurements were conducted using the Dektak XT profilometer with a tip diameter of 2 µm. The presented profile traverses (from bottom to top) from ITO to glass, with a measurement taken after each layer deposition. The height profile on the right is obtained by averaging across the entire width of the measurement.*

---

t- top layer, b- bottom layer

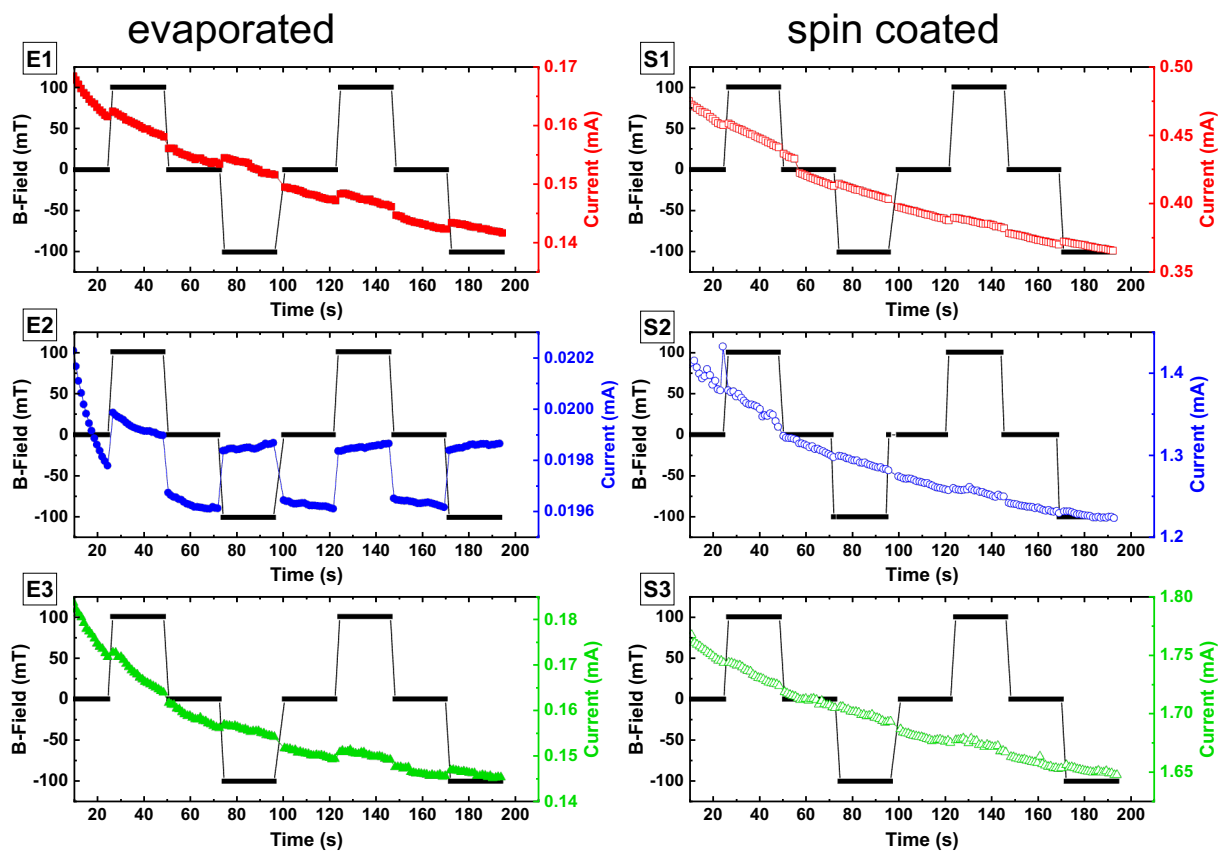

*Fig. S 3* Raw data of the variation of the current of the different OLEDs (E1: PVK; TCTA; TPBi (evaporated), E2: PVK:TCTA (mixed 1:1.5); TPBi (evaporated), E3: TCTA; TPBi (evaporated); S1: PVK; TCTA; TPBi (spin-coated), S2: PVK:TCTA (mixed 1:1.5); TPBi (spin-coated), S3: TCTA; TPBi (spin-coated) as a function of magnetic field alternatingly switched on (B-field = 100mT), off (B-field = 0mT), and on (B-field = -100mT)

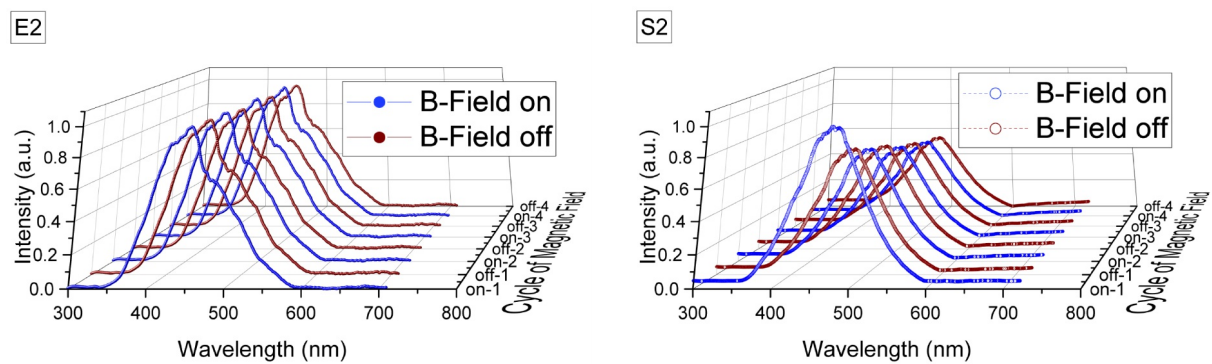

**Fig. S 4**. *Spectra recorded at each on-off cycle for the devices E2 and S2, respectively (cf. Figure 4. The external magnetic field was alternately switched on and off ($B_{on}$ = 100 mT) and the spectra were recorded within the first five seconds of each cycle. The device S2 reveals a much stronger luminescence decrease over time compared to the evaporated counterpart.*

*Table S 1.* Comparison of individual peak maxima and FWHMs during the application (±100 mT) and switching off (0 mT) of the magnetic field. The calculations are based on a nonlinear curve fit, where a Gaussian curve was fitted to the spectrum to each peak. The presented values represent the mean values of the respective on and off cycles

| Device | E1 | | E2 | | E3 | | S1 | | S2 | | S3 | |
|---|---|---|---|---|---|---|---|---|---|---|---|---|
| Magnetic field | on | off | on | off | on | off | on | off | on | off | on | off |
| Gauss-fit maxima [nm] | 459.51 ±0.05 | 459.64 ±0.08 | 450.27 ±0.03 | 450.33 ±0.06 | 461.57 ±0.06 | 470.31 ±0.99 | 470.59 ±0.52 | 470.31 ±0.99 | 474.86 ±1.21 | 475.19 ±0.49 | 473.30 ±0.96 | 474.04 ±0.73 |
| FWHM [nm] | 92.76 ±0.27 | 92.64 ±0.47 | 109.78 ±0.05 | 109.69 ±0.04 | 101.87 ±0.10 | 99.95 ±0.80 | 99.85 ±0.61 | 99.95 ±0.80 | 99.39 ±1.21 | 99.14 ±0.73 | 102.21 ±0.17 | 101.89 ±0.40 |

*Table S 2.* Maximum OMC values extracted from the diagrams in Fig. 3.

| Device | $OMC_{max}$ / % |
|---|---|
| S1 | (0.94 ± 0.10) |
| <span style="color:red">S2</span> | <span style="color:red">(2.54 ± 0.57)</span> |
| S3 | (0.12± 0.24) |
| E1 | (0.96± 0.05) |
| E2 | (1.32± 0.05) |
| E3 | (1.04± 0.09) |